\documentclass[conference]{IEEEtran}
\IEEEoverridecommandlockouts
\usepackage{cite}
\usepackage{amsmath,amssymb,amsfonts}
\usepackage{algorithmic}
\usepackage{graphicx}
\usepackage{textcomp}
\usepackage{xcolor}
\usepackage{hyperref}
\def\BibTeX{{\rm B\kern-.05em{\sc i\kern-.025em b}\kern-.08em
    T\kern-.1667em\lower.7ex\hbox{E}\kern-.125emX}}
\begin{document}

\title{SlicerChat: Building a Local Chatbot for 3D Slicer}

\author{\IEEEauthorblockN{Colton Barr}
\IEEEauthorblockA{\textit{Laboratory for Percutaneous Surgery, Queen's University, Queen's University}\\
\textit{Golby Lab, Brigham and Women's Hospital, Harvard Medical School}\\
c.barr@queensu.ca}

}

\maketitle

\begin{abstract}
3D Slicer is a powerful platform for 3D data visualization and analysis, but has a significant learning curve for new users. Generative AI applications, such as ChatGPT, have emerged as a potential method of bridging the gap between various sources of documentation using natural language. The limited exposure of LLM services to 3D Slicer documentation, however, means that ChatGPT and related services tend to suffer from significant hallucination. The objective of this project is to build a chatbot architecture, called SlicerChat, that is optimized to answer 3D Slicer related questions and able to run locally using an open-source model. The core research questions explored in this work revolve around the answer quality and speed differences due to fine-tuning, model size, and the type of domain knowledge included in the prompt. A prototype SlicerChat system was built as a custom extension in 3D Slicer based on the Code-Llama Instruct architecture. Models of size 1.1B, 7B and 13B were fine-tuned using Low rank Adaptation, and various sources of 3D Slicer documentation were compiled for use in a Retrieval Augmented Generation paradigm. Testing various combinations of fine-tuning and model sizes on a benchmark dataset of five 3D Slicer questions revealed that fine-tuning had no impact on model performance or speed compared to the base architecture, and that larger models performed slightly better with a significant speed decrease. Experiments with adding 3D Slicer documentation to the prompt showed that Python sample code and Markdown documentation were the most useful information to include in the prompt, but that adding 3D Slicer scene data and questions taken from Discourse also improved model performance. In conclusion, this project shows the potential for integrating a high quality, local chatbot directly into 3D Slicer to help new users and experienced developers alike to more efficiently use the software. Future work will focus on further testing with more reviewers and a larger benchmark dataset, as well as refining the fine-tuning approach and distilling down the retrieval augmented generation knowledge-base.
\end{abstract}

\section{Introduction}
3D Slicer refers to an ecosystem of data acquisition, processing and visualization tools built around an open-source software model. Originally created as a Master’s thesis project at MIT in 1998, 3D Slicer has since been downloaded by over one million users and maintained through over \$50 million dollars of research funding \cite{fedorov20123d}. The core 3D Slicer software library is built around a series of smaller, more broadly applicable software packages, including the Insight Tool Kit (ITK) and Visualization Tool Kit (VTK), and is most frequently access through the 3D Slicer desktop application or using the Python programming language. The desktop application consists of a variety of built-in GUI-based apps, called extensions, as well as over 160+ officially supported 3rd party extensions that can be easily installed. It also has a built-in Python interactor, enabling users to directly access all components of the 3D Slicer ecosystem using Python within the application. To support this complex amalgam of user interface functionality, Python packages, built-in extensions and existing projects within 3D Slicer, there are a variety of online resources for documentation and guidance. These resources vary from high level tutorials generated by a variety of different institutions, basic Quick Start guides within the official 3D Slicer documentation, in-depth user interface documentation, developer-oriented Python and C++ documents, as well as auto-generated ReadTheDocs sites for the C++ components, detailed extension documentation, and a highly active user forum on Discourse. The complexity of accessing C++ based VTK and ITK tools from Python also means that simple tab-completion of method names within the Python interactor is an important resource for otherwise undocumented methods.

A major challenge encountered by new users of 3D Slicer is knowing where to find all these resources and which resources to consult for different sources of information.  This frequently leads to users failing to take full advantage of the extensive existing tools built into the platform and re-implementing them, or else using the Discourse forum to ask questions that are answered in documentation. The most significant disruption of the 3D Slicer documentation “landscape” came in November of 2022 when ChatGPT was released, and subsequently in the spring of 2023 when it gained widespread traction. The ability of ChatGPT to quickly summarize and provide advice for previously obscure, poorly documented methods in 3D Slicer made it a popular resource and subject of frequent discussion on the 3D Slicer Discourse forum. The well-known limitations of large language models (LLMs), however, particularly for knowledge-intensive tasks like 3D Slicer question and answering, has led to mixed recommendations for the use of the tool among core 3D Slicer developers. The major challenge in the context of 3D Slicer Python development is hallucination, the tendency for LLMs to fabricate plausible but non-existant methods and functionalities. Additional concerns surrounding data and intellectual property privacy, out-of-date training data, and subscription costs have hampered the potential impact of ChatGPT and other generative AI models on the 3D Slicer community.

In contrast to the closed-source LLMs services that have emerged from major players like OpenAI, Microsoft and Google, there has been an accelerating push for competitive open-source LLMs that can be trained, tuned and run locally. Meta AI has led the field of open-source foundation models, with the release of Llama in February 2023 and Llama 2 in July, and other organizations like Mistral are following suit with their own highly capable open-source releases \cite{touvron2023llama} \cite{jiang2023mistral}. With the trend of higher quality, lower parameter open-source models, combined with the constant uptick in consumer compute capabilities, open-source LLMs are becoming an attractive alternative to using services like ChatGPT. A fundamental advantage of using open-source models is their extensibility: the direct access to model weights makes training, fine-tuning, and any number of prompt fine-tuning strategies potentially viable for optimizing a model to a particular use-case. For example, the Low Rank Adaptation method for parameter efficient LLM fine-tuning significantly reduces the compute and data required for tuning \cite{hu2021lora}. This vastly reduces the barriers to custom tuning of existing open-source LLMs, opening the possibility of producing better specialized LLMs for a particular task. For knowledge-intensive tasks, prompt engineering strategies such as injecting relevant base information through Retrieval Augmented Generation (RAG) or giving the model example outputs through few-shot prompting can further improve model performance \cite{wang2020generalizing} \cite{lewis2020retrieval}. These strategies, which are discussed further in the Related Works section, are leveraged in this project to build a chatbot specialized in answering 3D Slicer related questions. 

The objective of this project is to develop and integrate a local chatbot into 3D Slicer that has been optimized for answering Slicer-related questions. The chatbot, referred to here as SlicerChat (\href{https://github.com/ColtonBarr/SlicerChat}{https://github.com/ColtonBarr/SlicerChat}), will aim to cover the full breadth of questions a user of 3D Slicer might have, from high level questions about the user interface to technical API related requests, with a particular focus on servicing users that are using the Python interpreter to interact with Slicer. This project is meant to serve as a major step towards building an extension into the Extension Manager of 3D Slicer that would be easily accessible for both beginner users and experienced 3D Slicer developers. The focus of this project will be on methods of optimizing SlicerChat to use existing sources of 3D Slicer documentation most effectively to minimize hallucination and maximize the quality of the chatbot outputs.

\section{Research Questions}

There are two specific research questions, core to the eventual deployment and practical use of a 3D Slicer chatbot, that were used to motivate this work. The first research question is how model size and fine-tuning influence the quality and speed of answers generated by the chatbot. This RQ was selected to address a critical limitation of local LLMs: compute. In asking a user to deploy an LLM locally on their machine, the compute available for each user places a performance ceiling on their particular instance of the chatbot. In general, a larger number of parameters for an LLM tends to be associated with higher performance on benchmark tasks, while requiring more compute and potentially decreasing the speed of the resulting output, or in other words, increasing inference time. The inverse is true as well, with smaller models showing poorer performance on benchmark tasks but imposing a lower compute requirement and showing faster inference times. Understanding the impact of model size on the quality and speed of output for a 3D Slicer chatbot is critical for establishing bounds on performance and compute requirements for the application. Comparing these base models with fine-tuned models serves to clarify the utility and expected performance gains from fine-tuning, and further motivates future experiments to optimize the Slicer chatbot.

The second research question this project aims to answer is which 3D Slicer documentation is most useful in a Retrieval Augmented Generation (RAG) pipeline. The various forms and sources of 3D Slicer documentation all tend to encode separate but related information. Determining which sources of documentation are best suited for the RAG paradigm is a challenging task that requires a combination of domain knowledge and experimentation. The limited context-window of LLMs and increased inference time with larger prompts also imposes a requirement to provide concise, highly relevant information to the RAG pipeline. Giving the chatbot access to the same core documentation sources that an experienced 3D Slicer user would consult, and experimentally determining which knowledge is most useful during RAG, will help to reduce less useful information in the knowledge store and simplify the RAG process. This research question will use similar metrics of inference time and answer quality to evaluate the utility of different sources of knowledge for use during RAG.

\section{Methods}

\subsection{Data}
Selecting the relevant data for our chatbot was a core step for both the fine-tuning and RAG steps of the project. Through consultation with 3D Slicer experts and the chief architects of the platform, the most information-rich documentation sources were determined to be the Discourse forum, the 3D Slicer core Markdown documentation, and Python sample code from various sources. These latter two resources were obtained by downloading the core 3D Slicer remote repository from GitHub (\href{https://github.com/Slicer/Slicer}{https://github.com/Slicer/Slicer}), and extracting all Markdown (.md) files and Python (.py) files based on their file extensions. It is important to note that Python is used both as a scripting language to quickly interact with the 3D Slicer library, as well as a compiled language for building 3D Slicer extensions. The officially supported, but typically user-contributed, extensions available in the Extension Manager in 3D Slicer tend to be implemented in Python and are therefore an excellent source of sample 3D Slicer Python code. Since an important component of using Slicer efficiently is understanding what packages and functionalities have already been implemented, scraping the full corpus of Python and Markdown files from all 160 3D Slicer Extensions serves to both broaden the knowledge of the model as well as provide high quality examples of Python code. 

Extracting useful data from the 3D Slicer Discourse forum presented more of a challenge, since questions are used to start new threads that can be up to 100 posts long and include a variety of sub-questions and answers. To distill the most helpful information from Discourse, and create a supervised question-and-answer dataset that could be used for training and validation, only answers generated by the top 10 Discourse contributors and officially accepted by the original poster were scraped from the website. This resulted in a dataset of 2048 question-and-answer pairs. Another source of data used during RAG was information about the current state of the 3D Slicer desktop application at the time of the question being posed. This data was extracted from 3D Slicer by converting the overall status of the application, the Medical Reality Markup Language (MRML) Scene, to an XML string. From this XML string, the information in some highly informative tags was extracted then compressed, before being passed to the LLM for use in a RAG pipeline. This will be explored further in the sections that follow.

\begin{figure*}[h]
\centering
\includegraphics[width=0.9\textwidth]{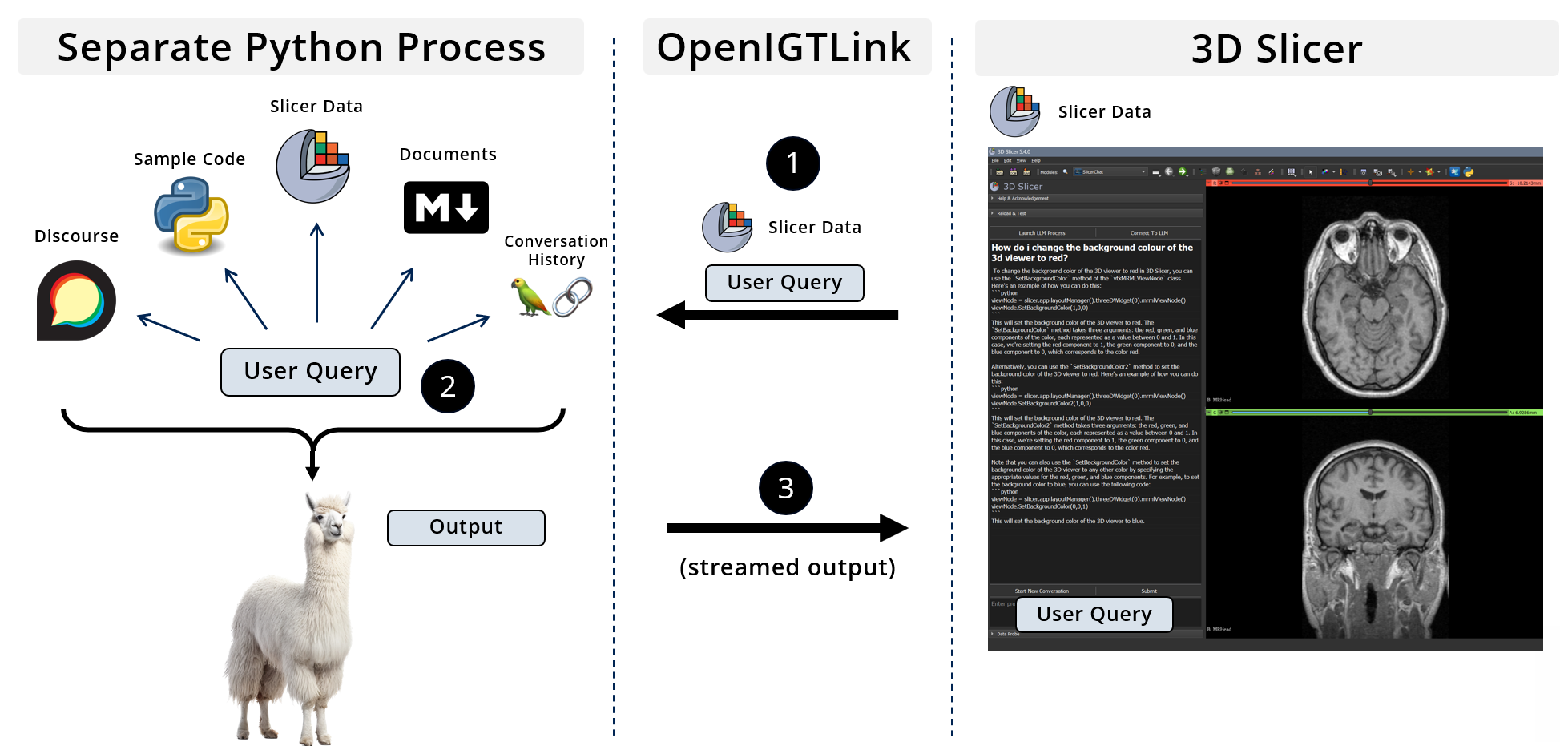}
\caption{The complete pipeline of SlicerChat. The user starts by entering a prompt, and the extension extracts the current 3D Slicer scene data. This data is then passed along with the query to the seperate python process, where it is accessed as one of 5 sources of RAG data. The resulting RAG prompt is passed to the selected model, and the output from the model is streamed back to 3D Slicer.}
\label{fig:ArchitectureDiagram}
\end{figure*}

\subsection{Model Selection and Finetuning}

The models used in this project were selected primarily for their state-of-the-art performance among open models on well-cited benchmarks including Human Eval and MBPP at the time of release \cite{roziere2023code}. The availability of different model sizes within the same Code Llama family of models made it an attractive choice for testing the effect of using different parameter sizes. In addition, their impressive capabilities with both programming tasks and general language reasoning makes them well suited to the potentially large variety of different types of questions they would receive from users. The specific models selected were CodeLlama-Instruct-13B-GPTQ, CodeLlama-Instruct-7B-GPTQ, and TinyLlama-1.1B-Chat-v0.3-GPTQ. The sizes of models were selected around the requirement of being able to fit onto a consumer grade laptop GPU, specifically an RTX 3070 8GB. The 13B model requires use of virtual VRAM to perform inference, but this is a reasonable compromise to run a much larger model. 

For fine-tuning these models, the Low Rank Adaptation (LoRA) method was selected and run using a 24GB desktop RTX 3090TI. The 2048 question-and-answer pairs obtained from Discourse were used for fine-tuning, and Bayesian hyperparameter searches were performed for all models. The specific tuning details are covered in the Experiments section.

\begin{figure}[!htb]
\centering
\includegraphics[width=0.489\textwidth]{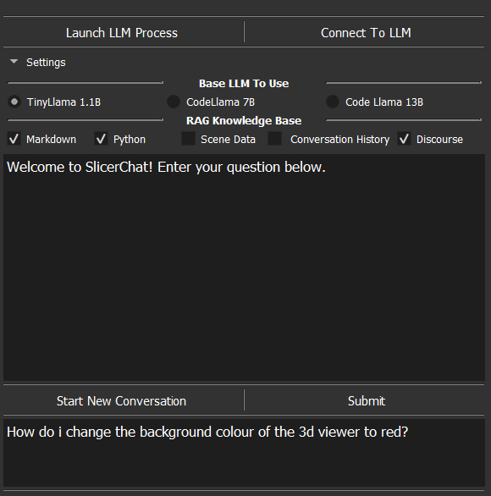}
\caption{The 3D Slicer extension UI for SlicerChat. It includes buttons for starting and connecting to the external LLM process, configuring the base LLM and RAG knowledge to include in the prompt, as well as resetting the conversation and submitting th input prompt. The larger dark square contains the streamed output tokens, while the smaller rectangle at the bottom is where the user enters their prompt.}
\label{fig:chatUI}
\end{figure}

\subsection{Retrieval Augmented Generation}

There were several important design decisions made during the implementation of RAG. The first was how to arrange the sources of knowledge into vector databases for matching to queries at inference time. Following collection of all Python and Markdown files from over $160$ repositories related to 3D Slicer core and various user-generated extensions, a total of $1.5$ million lines of Python and $500,000$ lines of Markdown were gathered into large numpy files. The decision was made to split the Python sample code and Markdown documentation into separate vector stores to force the RAG pipeline to find Python examples in addition to relevant documents for each query. Text splitters specifically designed for Python and Markdown were used from the LangChain package to chunk the text into 200 token chunks with a 50 token chunk overlap. These chunks were then embedded using the all-MiniLM-L6-v2 sentence transformer and stored in a Facebook AI Similarity Search (FAISS) vector store locally \cite{johnson2019billion}. For the Discourse question-answer pairs, the full questions and answers were embedded as a single chunk into their own FAISS vector store. This decision was made to facilitate the fetching of questions and answers with relevant language to enable the integration of 1-shot prompting using a relevant Discourse question. The chunking, embedding and generation of vector stores was performed offline for the Python, Markdown and Discourse databases, such that the pre-generated FAISS vector stores are loaded when the model is started by the user. 

Extracting relevant information from the current 3D Slicer scene to include in the RAG prompt posed several challenges, including the intelligent selection of relevant scene components and integration with the RAG pipeline. After converting the current 3D Slicer MRML scene to an .XML string, the most relevant XML tags were deemed to be those that contained the high level descriptions of all object types and names within the current scene. This data is then converted to a JSON string and passed to the Python thread containing the model, converted to a FAISS vector store, and queried using the user input to find the most relevant chunk of information in the 3D Slicer scene. To facilitate conversation and follow-up questions, the conversation history is also chunked and written to a FAISS vector store following each reply from the model. This is intended to facilitate a continuous chain of conversation without overwhelming the model’s context window with increasingly large prompts. While this functionality is necessary to create a true chatbot for 3D Slicer, the conversation history functionality was turned off during this project and need to be tested further in future studies.

\subsection{3D Slicer Extension and Architecture}

A unique opportunity with using local LLMs, particularly in the highly customizable development environment of 3D Slicer, is implementing the user interface directly within the application it is intended to service. In the case of 3D Slicer, building the SlicerChat interface as a custom extension such that it appears in the same menus and location as the other core functionalities of the software is aimed at to streamlining the user experience and simplifying the process of accessing knowledge from documentation. Implementation of large AI models within 3D Slicer, particularly through Python, tends to be complicated by the Global Interpreter Lock (GIL) that 3D Slicer enforces. The GIL makes true multithreading using Python in 3D Slicer infeasible, and results in all aspects of 3D Slicer processing hanging until a large computation, such as LLM inference, has completed execution. To overcome this limitation, previous projects have implemented a local server to pass data to a separate Python thread on the same device, generate the output, and stream this output back to 3D Slicer \cite{wasserthal2023totalsegmentator} \cite{liu2023samm}. A similar approach was taken in this project using the OpenIGTLink extension within 3D Slicer and the pyIGTL package in Python to setup a local server-client connection with an external python process \cite{tokuda2009openigtlink}. 

To facilitate this, the custom 3D Slicer extension starts the launching an external Python process. Shown in Figure \ref{fig:chatUI}, this extension can be used to launch and connect to the separate Python process containing the LLM. It also has menu options for selecting which Base LLM to use, specifying which RAG knowledge to include in the prompt, as well as start a new conversation and submit the prompt. The prompt and results appear in Qt text windows well suited for rich text and HTML-based formatting. Once the user hits the Submit button, the 3D Slicer extension immediately extracts the relevant XML tags in the current 3D Slicer scene. This data, along with the prompt and the selected RAG knowledge components, is compressed into a JSON string and passed to the parallel Conda process through OpenIGTLink. The prompt is then used to query the FAISS vector stores containing Python data, Markdown data, Discourse data, 3D Slicer Scene data, and the current conversation history, respectively. The generated RAG prompt is pass to the loaded model and the streaming interface within LangChain is used to facilitate token-by-token streaming of the output over PyIGTL back to the 3D Slicer extension. The complete step-by-step process of receiving a prompt and generating an output is shown in Figure \ref{fig:ArchitectureDiagram}.

\section{Experiments}

\subsection{Benchmark Dataset}

\begin{figure}[!htb]
\centering
\includegraphics[width=0.3\textwidth]{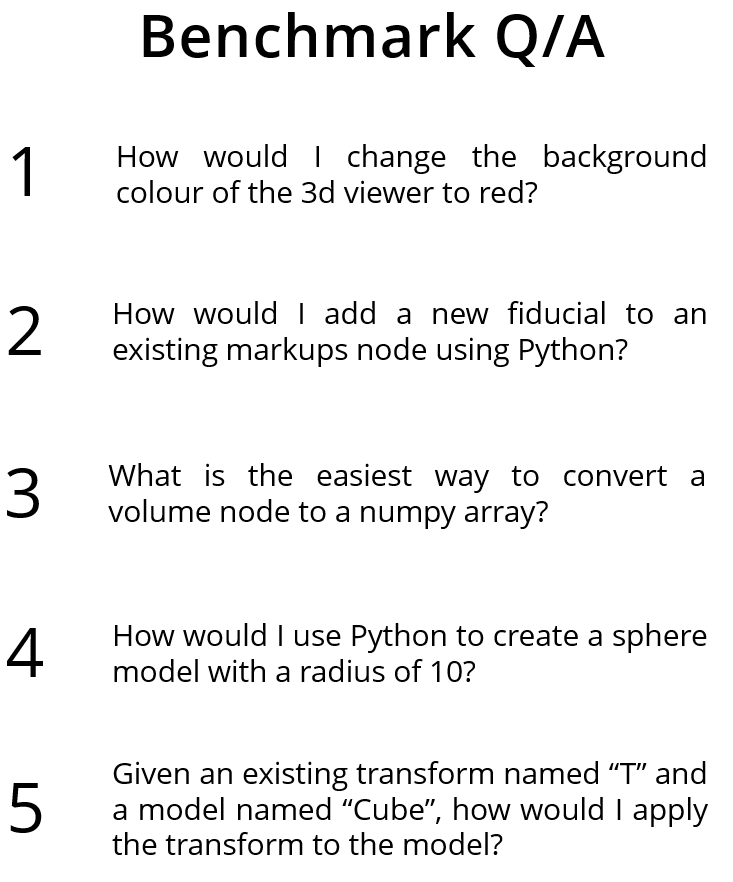}
\caption{The model performance on each of the 5 benchmark questions, grouped by the RAG knowledge made available at inference time.}
\label{fig:BenchmarkQs}
\end{figure}

In an effort to answer the research questions posed for this project, a series of experiments were conducted. These experiments revolved around the use of a benchmark question-and-answer dataset to evaluate the inference speed and output quality of different model architectures, both fine-tuned and base models, across a set of different RAG prompting regimes. To avoid the increasing challenges associated with data leakage and testing on previously seen training data, the questions in this benchmark dataset were generated by hand in collaboration with 3D Slicer architects and experts. They reflect common challenges faced by beginner to intermediate users, and the expected answers to each must include both natural language and testable Python code. They are also unique in that there tends to be a highly efficient solution in the form of a poorly documented function call that solves each of them. The complete list of benchmark questions are shown in Figure \ref{fig:BenchmarkQs}.

Given the complex combination of Python code and natural language output necessary to answer 3D Slicer related questions, the decision was made to evaluate the network outputs manually by an expert 3D Slicer reviewer. Due to time constraints, a single reviewer was used without implementing blinding or pairwise comparison. The reviewer instead attempted to directly run the Python code in the 3D Slicer Python interpreter, and scored the model output based on the number of lines of code that ran successfully. A complete solution that ran without modification received a 5 out of 5, while a model output for which only 20\% of the code ran for received a 1 out of 5. Comments were also recorded to note any unique or unexpected behaviours of the networks. For each model output, the execution time was also recorded as the time between sending the prompt from 3D Slicer and receiving an End Of String (EOS) token from the LLM. 

\subsection{RQ1: Comparison of Model Size and Fine-tuning}

The first research question posed in this work was the effect of model size and model fine-tuning on the performance of a given LLM. To adequately test this, the three selected models were fine-tuned on the full set of 2048 Discourse questions and answers using LoRA. To maintain consistency across the model tuning attempts, a similar Bayesian hyperparameter tuning was performed for all three networks using the Weights and Biases service. The hyperparameters that were optimized include Lora alpha, dropout and rank, gradient accumulation steps, learning rate, optimizer, training batch size and the number of warmup steps. All networks were trained for 300 steps, with checkpoints saved every 30 steps, and the best performing network from each was used at its checkpoint with the lowest loss on the validation set. The fine-tuned networks were then integrated into the 3D Slicer extension and tested using a complete set of RAG data for all 5 questions in the benchmark dataset. This test was followed by the same evaluation of the base models for each architecture size. The reported metrics for this experiment include the reviewer rating for each question output as well as the inference time for each question.

\subsection{RQ2: Comparison of RAG Knowledge Sources}

To better understand the importance of different RAG data sources on the quality of the model output, a separate experiment was conducted to answer the second research question. Using a base model of the 7 billion parameter CodeLlama architecture, several combinations of RAG data were passed to the model at inference time and the resulting model performance was recorded by the reviewer. The separate groups of RAG data that were tested include: a combination of Python and Markdown data; strictly 3D Slicer scene data; 1-shot prompting using Discourse data; and a combination of Python, Markdown, 3D Slicer data and Discourse data. Across these four combination of RAG data, the set of 5 questions in the benchmark dataset was posed, and the human reviewer output was recorded for each question. 

\section{Results}

\subsection{Model Fine-Tuning}

\begin{figure*}[h]
\centering
\includegraphics[width=1\textwidth]{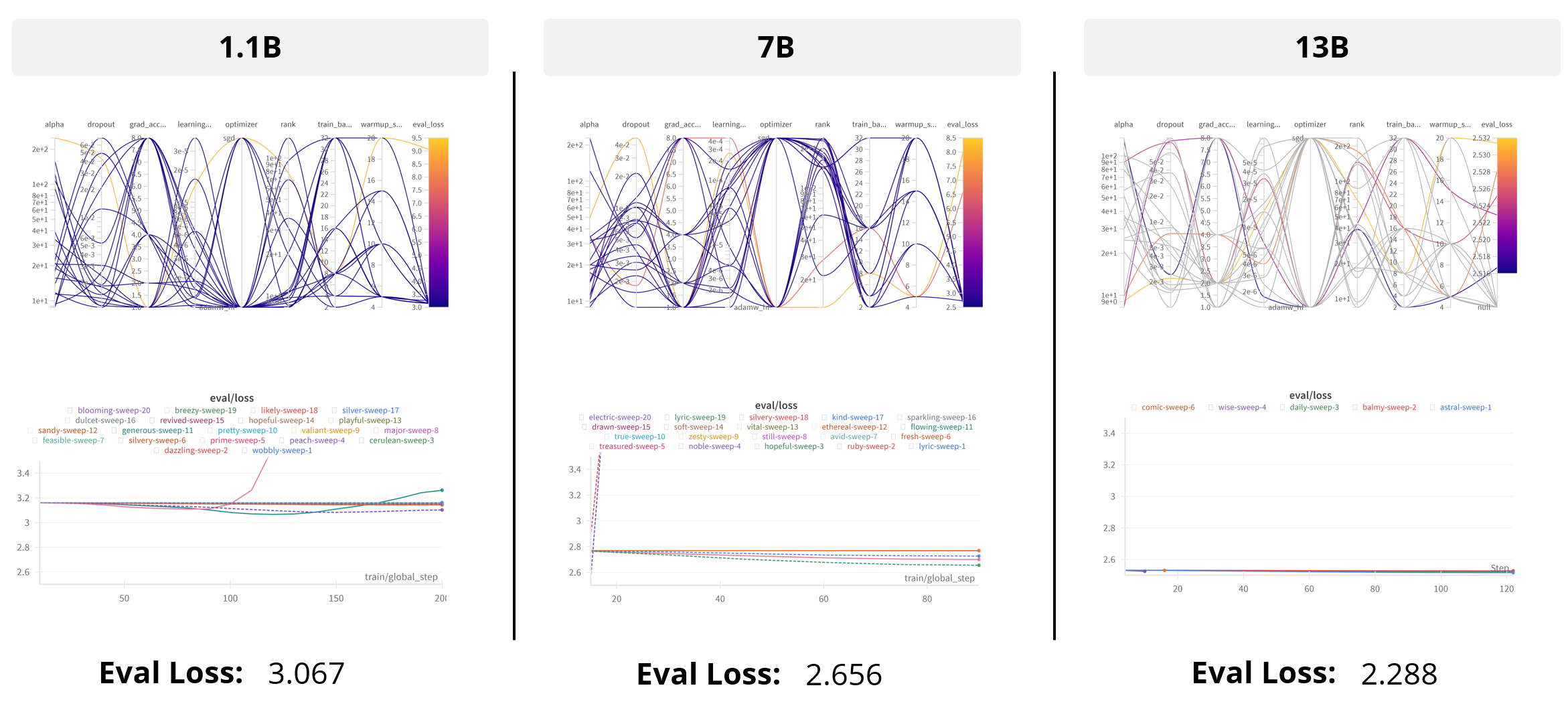}
\caption{The fine-tuning outputs for all three model architecture sizes. The first row contains the results of the Bayesian Hyperparameter sweeps using Weights and Biases, while the second row shows the loss curve on the evaluation set for each network tested, and the final row indicates the lowest evaluation loss value obtained for all tested networks.}
\label{fig:FinetuneResults}
\end{figure*}

Bayesian hyperparameter optimization was performed using Weights and Biases for $20$ iterations on the 1.1B, 7B and 13B parameter models. Figure \ref{fig:FinetuneResults} shows the hyperparameter optimization output and validation set loss curves. The best performing 1.1B parameter network achieved a lowest validation loss of $3.067$, compared to the best 7B network showing a loss of $2.656$ and the best performing 13B network showing a loss of $2.288$ on the validation set. Note that hardware limitations reduces the number of successful model training runs using the 13B network to only 4, and subsequent single network fine-tuning runs were used to achieve the validation set loss of $2.288$.

\subsection{Comparison of Model Fine-tuning and Architecture}

\begin{figure}[!htb]
\centering
\includegraphics[width=0.5\textwidth]{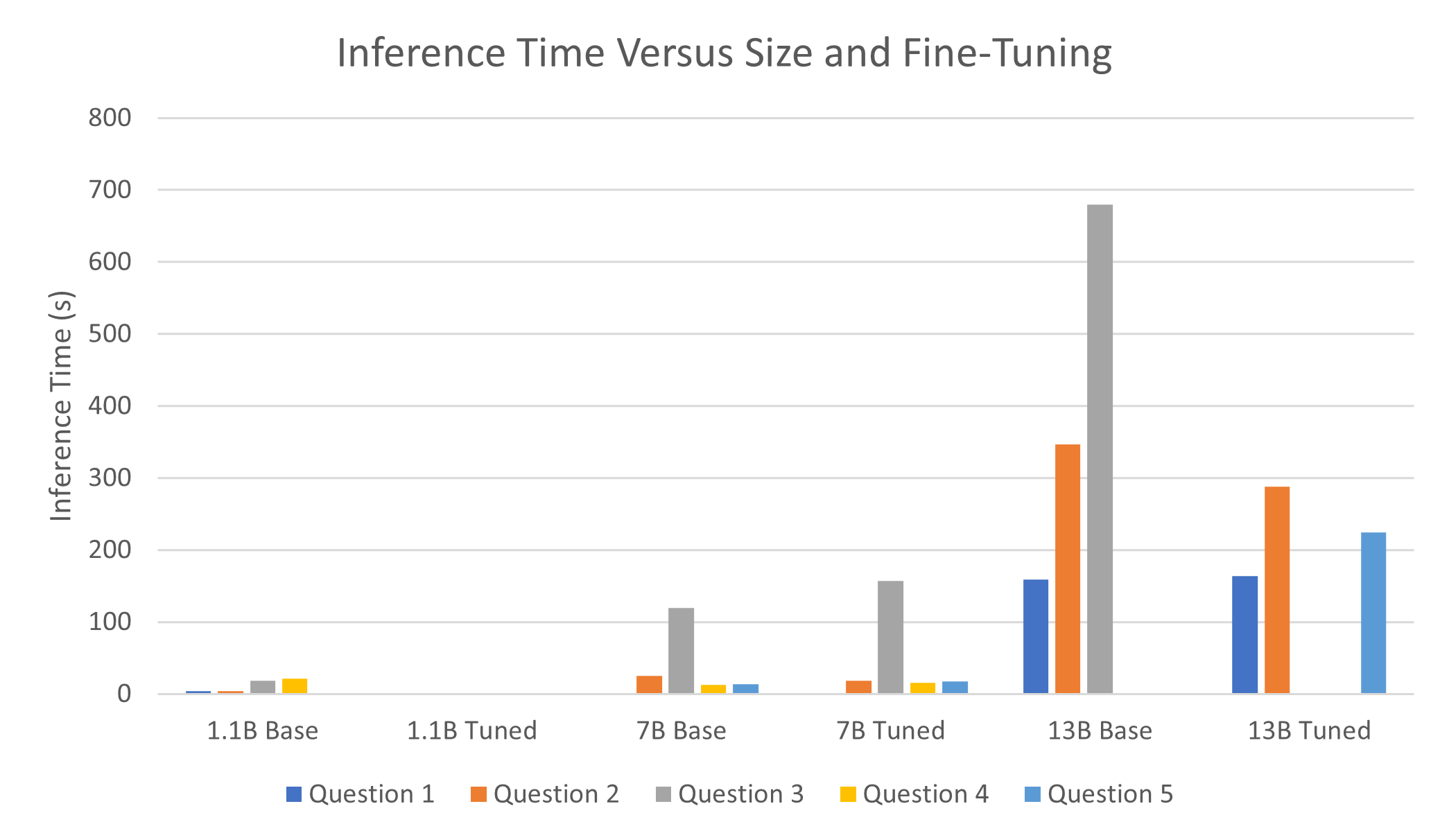}
\caption{The inference time for each question in seconds grouped by the model used to generate the inference time. Note that for each model size of either 1.1B, 7B or 13B, the base model and fine-tuned models are tested.}
\label{fig:RQ1InferenceTime}
\end{figure}

\begin{figure}[!htb]
\centering
\includegraphics[width=0.5\textwidth]{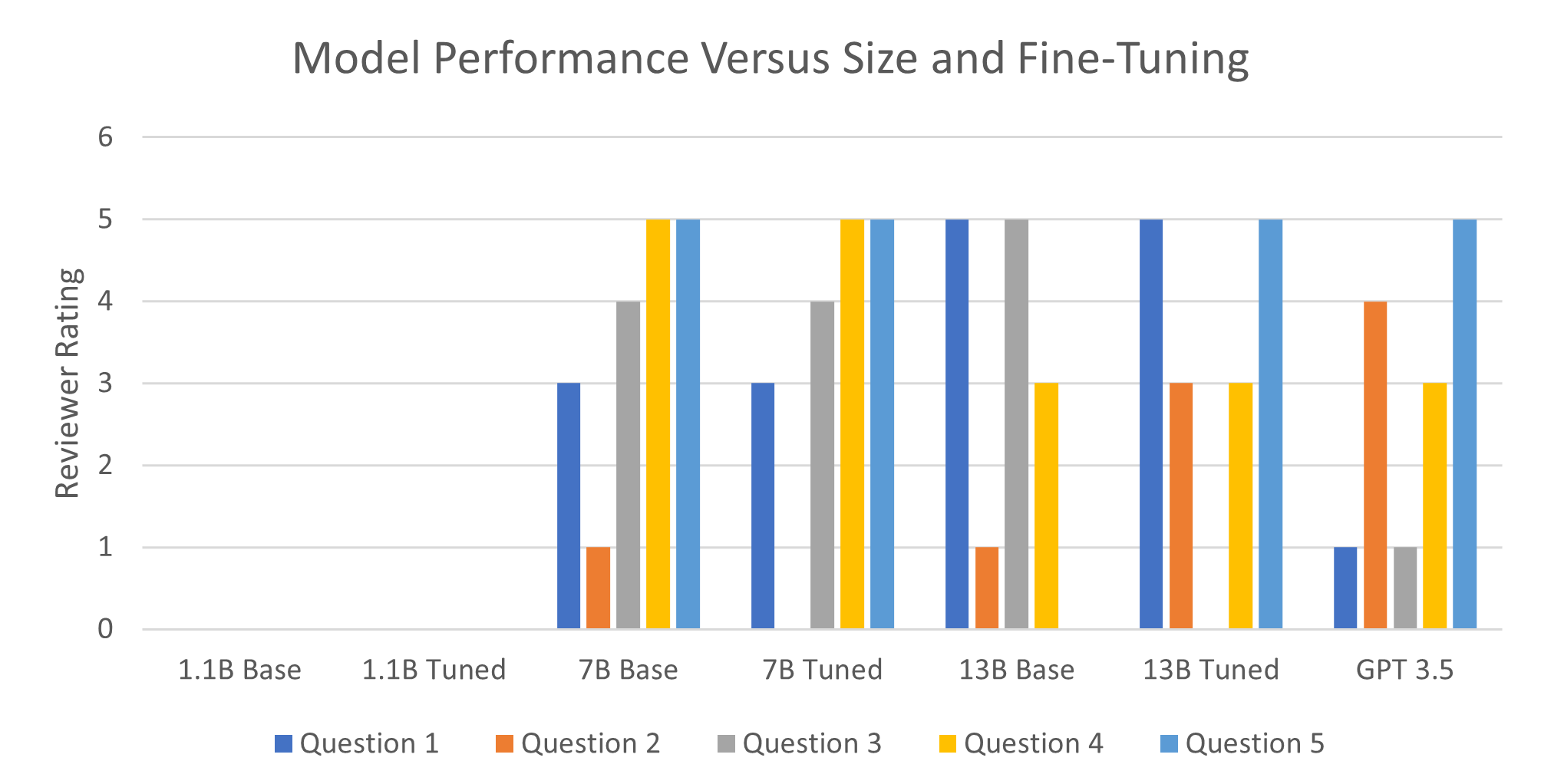}
\caption{The model performance on each of the 5 benchmark questions, grouped by each combination of parameter size and fine-tuning.}
\label{fig:RQ1ModelPerformance}
\end{figure}

The results of using the best performing fine-tuned networks of each parameter size to answer the benchmark questions are shown in Figures \ref{fig:RQ1InferenceTime} and \ref{fig:RQ1ModelPerformance}, alongside the performance of the base model of each architecture. Figure \ref{fig:RQ1InferenceTime} shows the reviewer rating between 0 and 5 for each question, grouped by the network that produced the output. Figure \ref{fig:RQ1ModelPerformance} shows the inference time in seconds for the same groupings of network type and question being answered. Note that GPT 3.5, accessed through the ChatGPT interface, was also evaluated on the benchmark dataset for comparison. The user reviews of the output are included in the “Performance” graph, but were excluded from the “Inference Time” graph due to their negligible (less than 5 seconds) values. 

The 1.1B parameter TinyLlama network was unable to generate a usable output on all benchmark questions, across both the fine-tuned and base model architectures. These invalid outputs tended to be either repetition of the prompt, restating of the question, or Python code snippets that were not relevant to the input question. The 7B models were able to return meaningful outputs across both the fine-tuned and base models for all questions, with the exception of an invalid output for Question 2 using the tuned network. The 13B models exceeded the memory capabilities of the laptop GPU on two occasions: Question 5 using the base model and Question 3 using the tuned model. On Question 4 using the 13B model, an infinite loop was encountered that resulted in a non-meaningful inference time, but an output that still had some valid code within it. In this case the inference time is shown as zero, but the model performance is still evaluated based on the quality of the output at the time that the execution was halted. It is important to note that results were obtained by passing the models all relevant Python, Markdown, and Discourse data, as well as relevant 3D Slicer scene data generated through RAG, within their prompts at inference time. Conversation history was not included in these tests.

\subsection{Comparison of RAG Knowledge Sources}

\begin{figure}[!htb]
\centering
\includegraphics[width=0.5\textwidth]{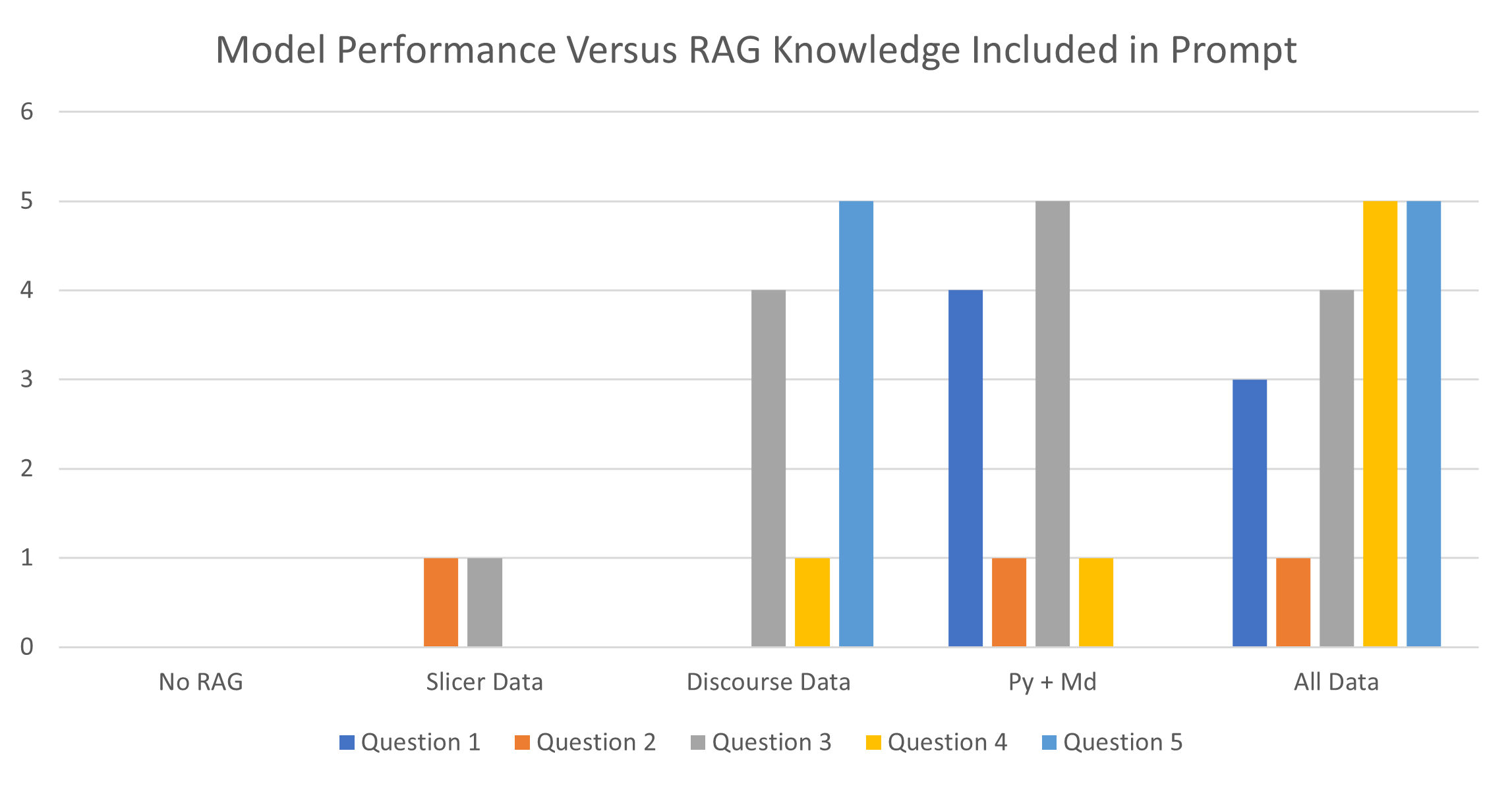}
\caption{The model performance on each of the 5 benchmark questions, grouped by the RAG knowledge made available at inference time.}
\label{fig:RQ2ModelPerformance}
\end{figure}

The results for the comparison of RAG knowledge sources are shown in Figure \ref{fig:RQ2ModelPerformance}, with the bar graphs for Questions 1 through 5 grouped according to the set of RAG data included in their prompts. All results are generated using the same Base 7B Code Llama Instruct model without fine-tuning. In this case none of the questions resulted in an infinite loop output, and all questions for which there is no visible bar produced Python code that was unusable or invalid resulting in a 0 out of 5 reviewer score. 

The reviewer noted in two cases that the model generated a solution that was previously not known to the reviewer. The first was for Question 3 using Discourse Data, where the model generated a shortcut method call that was more efficient than the expected answer. The second unexpected output was for Question 4 using all available RAG data, where the model returned a call to a method nested 5 classes deep that removed the need to use an external extension to produce a model of a sphere.

\section{Related Work}

This project builds on research from a variety of different domains, including open-source LLMs, optimization of LLM performance through fine-tuning and prompt engineering, integration of AI into 3D Slicer, and the use of a chatbot for interacting with technical documentation. The specific open-source LLMs used in this project were CodeLlama-Instruct-13B-GPTQ, CodeLlama-Instruct-7B-GPTQ, and TinyLlama-1.1B-Chat-v0.3-GPTQ, all accessed from TheBloke on HuggingFace. All three of these networks were based around the Code Llama foundation models released by Meta AI in August 2023, which are themselves fine-tuned versions of the Llama 2 architecture \cite{touvron2023llama} \cite{roziere2023code}. These models are extensively optimized on programming tasks, particularly Python data, and the “Instruct” variety of these models were further tuned on question and answering tasks following technical training on programming. These specific models were all quantized by TheBloke using the one-shot weight quantization method GPTQ, proposed by Frantar et al. \cite{frantar2022gptq}. The 1.1B parameter network TinyLlama is an ongoing project by Zhang et al. to train a 1.1 billion parameter model on 3 trillion tokens \cite{tinyllama}. 

Improving the performance of open-source LLMs using fine-tuning and prompt engineering is a highly active area of research, with numerous potential methods of improving the base performance of models for a specific task. In this project, we explored fine-tuning as a method of improving the CodeLlama networks for 3D Slicer related question-and-answering. For this we used the Low Rank Adaptation (LoRA) method of fine-tuning proposed by Hu et al. in 2021 \cite{hu2021lora}. This method was selected for its simplicity of implementation using the Parameter-Efficient Fine-Tuning (PEFT) package, as well as the significant decrease in required hardware for tuning and storage of the resulting weights. For prompt engineering, the primary technique used was the Retrieval Augmentation Generation (RAG) methodology first published by Lewis et al. in 2020 \cite{lewis2020retrieval}. This technique involves augmenting prompts to include knowledge that pertains to the question and proposed an efficient method for doing this using encoded chunks of knowledge. This method was used in combination with few-shot prompting principles, such that the concepts behind RAG are used to find the most relevant question-answer pair from a database of examples and this question-answer sample was included in the prompt as a form of one-shot learning \cite{wang2020generalizing}. 

This project borrowed inspiration from several related but distinct projects related to documentation chatbots and integration of AI into 3D Slicer. The challenges of interfacing with a large AI model in 3D Slicer have been explored by several previous projects. This includes the Slicer TotalSegmentator extension, which enables users to run a deep learning network for segmentation of CT scans locally, and SlicerSAM, the first published extension to the author’s knowledge that interfaced between Slicer and a foundational model \cite{liu2023samm} \cite{wasserthal2023totalsegmentator}. While the code in both these GitHub repositories was valuable throughout this project to assist with interfacing between a local LLM and 3D Slicer, their implementations had several limitations that had to be overcome to make them viable for use with LLMs. 
Another related project was the OHIF-Bot, a custom OpenAI chatbot implemented using the Assistants API in ChatGPT by Alireza Sedghi (\href{https://chat.openai.com/g/g-gFuX0AKlq-ohif-bot}{OHIF Bot}). OHIF is a web-based application with similar use-cases to 3D Slicer, but very little cross-over between API calls and technical backend. This OHIF-Bot has seen significant traction on LinkedIn, but has the inherent limitation of requiring a paid subscription to OpenAI to use. There was also an example of a custom RAG-driven chatbot posted on the 3D Slicer Discourse forum in early 2023 by a prominent member of the 3D Slicer community Dr. Rudolf Bumm (\href{https://discourse.slicer.org/t/langchain-query-the-complete-3d-slicer-documentation-script-repository-and-faq-pdf-and-html-with-openai-llm/28746}{Relevant Discourse Post}). This implementation also relied on the OpenAI API to run, however, and only used a very limited subset of the 3D Slicer documentation for RAG as well as using a Gradio web interface. The work presented in this project extends these concepts by building the chatbot directly into 3D Slicer, exploring a larger variety of documentation, using a local open-source model, and fine-tuning the model on Discourse question-and-answer pairs.

\section{Discussion}

The purpose of this project is to evaluate the current implementation of a 3D Slicer chatbot and better understand how to continue optimizing its performance. In order to motivate this work with concrete goals, two research questions were posed to explore how different design decisions will eventually impact the user experience with this chatbot platform. 

The first research question aimed to clarify how fine-tuning and model size impacted execution time and quality. Through performing fine-tuning of different architecture sizes and comparing their outputs to their base model counterparts, the results showed very subtle differences between base model and fine-tuned model pairs. In the case of the 1.1B model this was negligible since all outputs were invalid and infinitely recurring. For the 7B model the performance was identical between base and fine-tuned version, with the only subtlety being the base model achieving 1 out of 5 on Question 2, compared to 0 out of 5 on Question 2 for the fine-tuned model. Interpreting the performance of the base and fine-tuned 13B models is made challenging by the memory errors encountered inconsistently during testing, which included Question 5 for the base model and Question 3 for the fine-tuned model. Excluding questions 3 and 5, the only difference between the two versions of the 13B model was a 3/5 for the tuned model on Question 3 in comparison with a 1/5 for the base model. The difference in inference time between base and tuned models was likewise almost negligible, suggesting that there was no notable performance gain attributable to the use of fine-tuning on any network size. This is also consistent with the poor training performance and loss curves obtained during fine-tuning, which suggested minimal learning occurring during training.

This first experiment was also aimed at better understanding the differences in inference time between different model sizes. As illustrated in Figure \ref{fig:RQ1InferenceTime}, particularly between the 7B and 13B model, the differences were stark. The execution times for Question 2 on the 7B model, for example, were 18 seconds for the base model and 25 seconds for the fine-tuned model. This is compared to 346 seconds and 288 seconds on the base 13B and fine-tuned 13B models, respectively. This over 10x increase in inference time was seen across almost all questions for which there were valid outputs from both models, highlighting the substantial impact of using larger model sizes on network performance.

The second research question, pertaining to the importance of distinct forms of knowledge within the generation of a RAG prompt, was explored in the second experiment. The contrast between the performance of the default 7B model without any RAG knowledge included in the prompt and with all the available knowledge in the prompt was stark. The poor performance of the base models highlights that, whether due to limited parameter size or due to the training regime used for CodeLlama-Instruct, it possesses minimal niche 3D Slicer knowledge in its base form. The increase in performance between the base and full RAG model provides a strong justification for the inclusion of RAG data to improve model performance. The inconsistent benefits across different sets of RAG data seems to suggest different sources of knowledge were better suited for distinct questions in the evaluation set. For example, Question 5 received a 5/5 score when the Discourse data was present in the prompt, however in the absence of Discourse data all other attempts to answer the question received a 0/5. A similar trend was seen for Question 1 and the Python and Markdown data, where the absence of Python and Markdown consistently resulted in a 0 out of 5 score. Question 4 had an interesting progression that seemed to suggest the importance of multiple sources of knowledge to solve the question. The score of 1 out of 5 for both the Discourse data and the Py / Md documentation showed poor quality answers, while combing all potential sources of data yielded a 5/5 ranking and an obscure method call that was previously unknown to the reviewer. There did seem to be a slight decrease in answer quality for Question 1 when the Discourse and Slicer data were added in the “All Data” trial, highlighting the potentially negative impact of have extraneous information in the prompt. 

\section{Limitations}

There are a number of important limitations with this work that must be noted and addressed in future studies. A major limitation is the use of a single unblinded reviewer to evaluate the model performance in a manner that not strictly objective. The familiarity of the reviewer with the work and hypothesized results for each network, combined with their unblinded evaluation and subjective interpretation of the natural language outputs, may have resulted in biased performance evaluations. Combined with the limited size of the benchmark dataset, the results reported here should not be taken as a rigorous test of model performance. The impact of this evaluation method was mitigated by including a more objective “percentage of Python code that works” component to the performance evaluation and using a reviewer highly familiar with 3D Slicer, but must be made more robust in future works. The size of the benchmark dataset must also be expanded in future studies to better approach the full breadth of potential questions 3D Slicer users might pose to the chatbot, and cover more high level natural language inquiries as well as lower level developer questions. A promising future direction for this work is the development of a standardize 3D Slicer benchmark question-and-answering dataset that can be used to guide these and other future efforts to build a chatbot into the system.

Another limitation of this work was the method of fine-tuning the network and the poor quality of the results. While there were some differences between the outputs of networks that were fine-tuned versus the base architectures, these differences were very small and did not meaningfully impact the resulting performance evaluations of the networks. This was likely due to limited learning occurring during the fine-tuning phase of the project based on the consistently flat validation loss curves. Reasons for this poor fine-tuning performance could be the quality of question-answer pairs extracted from Discourse being limited, or the previous inclusion of this Discourse data in the original unsupervised training set used for Llama 2 and Code Llama. Future exploration into fine-tuning of 3D Slicer chatbots should focus on building a custom, high quality training set that the model cannot have already seen, and that has been carefully catered to improve model performance on 3D Slicer tasks. One possibility discussed during this project was forcing the model to produce executable Python code that can then be compiled and objectively tested using a Python kernel. 

A final limitation is that this project is missing a thorough investigation of prompt engineering and exploration of the resulting RAG prompts before being passed to the network. Understanding how the FAISS vector stores are handling the knowledge they contain, and subsequently how this knowledge may be used by the network, may have helped to better explain the difference seen in performance between different RAG knowledge tests. One future improvement that has been posed is being more transparent about the RAG knowledge that the model obtains at inference time, and including “references” to the relevant documentation or website directly within the answer. This could help to overcome lingering challenges with hallucination, and guarantee that users will at least be able to access the source documentation even if the model misinterprets it. A major next step for this project will be optimizing the knowledge stores to distill only the most relevant information, and in doing so the addition of documentation links for each chunk of knowledge would be an important consideration.

\bibliographystyle{IEEEtran}
\bibliography{mybib}

\begin{thebibliography}{10}
\providecommand{\url}[1]{#1}
\csname url@samestyle\endcsname
\providecommand{\newblock}{\relax}
\providecommand{\bibinfo}[2]{#2}
\providecommand{\BIBentrySTDinterwordspacing}{\spaceskip=0pt\relax}
\providecommand{\BIBentryALTinterwordstretchfactor}{4}
\providecommand{\BIBentryALTinterwordspacing}{\spaceskip=\fontdimen2\font plus
\BIBentryALTinterwordstretchfactor\fontdimen3\font minus \fontdimen4\font\relax}
\providecommand{\BIBforeignlanguage}[2]{{%
\expandafter\ifx\csname l@#1\endcsname\relax
\typeout{** WARNING: IEEEtran.bst: No hyphenation pattern has been}%
\typeout{** loaded for the language `#1'. Using the pattern for}%
\typeout{** the default language instead.}%
\else
\language=\csname l@#1\endcsname
\fi
#2}}
\providecommand{\BIBdecl}{\relax}
\BIBdecl

\bibitem{fedorov20123d}
A.~Fedorov, R.~Beichel, J.~Kalpathy-Cramer, J.~Finet, J.-C. Fillion-Robin, S.~Pujol, C.~Bauer, D.~Jennings, F.~Fennessy, M.~Sonka \emph{et~al.}, ``3d slicer as an image computing platform for the quantitative imaging network,'' \emph{Magnetic resonance imaging}, vol.~30, no.~9, pp. 1323--1341, 2012.

\bibitem{touvron2023llama}
H.~Touvron, L.~Martin, K.~Stone, P.~Albert, A.~Almahairi, Y.~Babaei, N.~Bashlykov, S.~Batra, P.~Bhargava, S.~Bhosale \emph{et~al.}, ``Llama 2: Open foundation and fine-tuned chat models,'' \emph{arXiv preprint arXiv:2307.09288}, 2023.

\bibitem{jiang2023mistral}
A.~Q. Jiang, A.~Sablayrolles, A.~Mensch, C.~Bamford, D.~S. Chaplot, D.~d.~l. Casas, F.~Bressand, G.~Lengyel, G.~Lample, L.~Saulnier \emph{et~al.}, ``Mistral 7b,'' \emph{arXiv preprint arXiv:2310.06825}, 2023.

\bibitem{hu2021lora}
E.~J. Hu, Y.~Shen, P.~Wallis, Z.~Allen-Zhu, Y.~Li, S.~Wang, L.~Wang, and W.~Chen, ``Lora: Low-rank adaptation of large language models,'' \emph{arXiv preprint arXiv:2106.09685}, 2021.

\bibitem{wang2020generalizing}
Y.~Wang, Q.~Yao, J.~T. Kwok, and L.~M. Ni, ``Generalizing from a few examples: A survey on few-shot learning,'' \emph{ACM computing surveys (csur)}, vol.~53, no.~3, pp. 1--34, 2020.

\bibitem{lewis2020retrieval}
P.~Lewis, E.~Perez, A.~Piktus, F.~Petroni, V.~Karpukhin, N.~Goyal, H.~K{\"u}ttler, M.~Lewis, W.-t. Yih, T.~Rockt{\"a}schel \emph{et~al.}, ``Retrieval-augmented generation for knowledge-intensive nlp tasks,'' \emph{Advances in Neural Information Processing Systems}, vol.~33, pp. 9459--9474, 2020.

\bibitem{roziere2023code}
B.~Roziere, J.~Gehring, F.~Gloeckle, S.~Sootla, I.~Gat, X.~E. Tan, Y.~Adi, J.~Liu, T.~Remez, J.~Rapin \emph{et~al.}, ``Code llama: Open foundation models for code,'' \emph{arXiv preprint arXiv:2308.12950}, 2023.

\bibitem{johnson2019billion}
J.~Johnson, M.~Douze, and H.~J{\'e}gou, ``Billion-scale similarity search with {GPUs},'' \emph{IEEE Transactions on Big Data}, vol.~7, no.~3, pp. 535--547, 2019.

\bibitem{wasserthal2023totalsegmentator}
J.~Wasserthal, H.-C. Breit, M.~T. Meyer, M.~Pradella, D.~Hinck, A.~W. Sauter, T.~Heye, D.~T. Boll, J.~Cyriac, S.~Yang \emph{et~al.}, ``Totalsegmentator: Robust segmentation of 104 anatomic structures in ct images,'' \emph{Radiology: Artificial Intelligence}, vol.~5, no.~5, 2023.

\bibitem{liu2023samm}
Y.~Liu, J.~Zhang, Z.~She, A.~Kheradmand, and M.~Armand, ``Samm (segment any medical model): A 3d slicer integration to sam,'' \emph{arXiv preprint arXiv:2304.05622}, 2023.

\bibitem{tokuda2009openigtlink}
J.~Tokuda, G.~S. Fischer, X.~Papademetris, Z.~Yaniv, L.~Ibanez, P.~Cheng, H.~Liu, J.~Blevins, J.~Arata, A.~J. Golby \emph{et~al.}, ``Openigtlink: an open network protocol for image-guided therapy environment,'' \emph{The International Journal of Medical Robotics and Computer Assisted Surgery}, vol.~5, no.~4, pp. 423--434, 2009.

\bibitem{frantar2022gptq}
E.~Frantar, S.~Ashkboos, T.~Hoefler, and D.~Alistarh, ``Gptq: Accurate post-training quantization for generative pre-trained transformers,'' \emph{arXiv preprint arXiv:2210.17323}, 2022.

\bibitem{tinyllama}
\BIBentryALTinterwordspacing
T.~W. Peiyuan~Zhang, Guangtao~Zeng and W.~Lu. (2023, Sep) Tinyllama. [Online]. Available: \url{https://github.com/jzhang38/TinyLlama}
\BIBentrySTDinterwordspacing

\end{thebibliography}


\end{document}